\documentclass[%
 reprint,
 amsmath, amssymb,
 aps,
]{revtex4-1}
\usepackage{natbib}
\usepackage{graphicx}% Include figure files
\usepackage{float}
\usepackage{dcolumn}% Align table columns on decimal point
\usepackage{bm}% bold math
\begin{document}

%\preprint{APS/123-QED}

\title{Dynamical coupling between protein conformational fluctuation and hydration water : Heterogeneous dynamics of biological water}% Force line breaks with \\

\author{Sayantan Mondal}
\author{Saumyak Mukherjee}%
\author{Biman Bagchi}
\email{Email: bbagchi@sscu.iisc.ernet.in, profbiman@gmail.com}
\affiliation{Solid State and Structural Chemistry Unit\\ Indian Institute of Science, Bangalore - 560012, India}%

\date{\today}% It is always \today, today,

\begin{abstract}
We investigate dynamical coupling between water and amino acid side-chain residues in solvation dynamics by selecting residues often used as natural probes, namely tryptophan, tyrosine and histidine, located at different positions on protein surface and having various degrees of solvent exposure. Such differently placed residues are found to exhibit different timescales of relaxation. The total solvation response, as measured by the probe is decomposed in terms of its interactions with (i) protein core, (ii) side-chain atoms and (iii) water molecules. Significant anti cross-correlations among these contributions are observed as a result of side-chain assisted energy flow between protein core and hydration layer, which is important for the proper functionality of a protein. It is also observed that there are rotationally faster as well as slower water molecules than that of bulk solvent, which are considered to be responsible for the multitude of timescales that are observed in solvation dynamics. We also establish that slow solvation derives a significant contribution from protein side-chain fluctuations. When the motion of the protein side-chains is forcefully quenched, solvation either becomes faster or slower depending on the location of the probe.

\begin{description}
\item[Keywords]
Biological water, Solvation dynamics, Protein hydration layer, anti-correlation, Energy decomposition, conformational fluctuation.
\end{description}
\end{abstract}

\maketitle

%\tableofcontents

\section{\label{sec:level1}INTRODUCTION}
The water layer occupying the interface between a protein molecule and the bulk water is termed as ‘protein hydration layer’ and the water inhabiting the layer is termed as “biological water”.\cite{RN48,RN5,RN19,RN21,RN13,RN27,RN85,RN20,RN41,RN17,RN18} It plays an important role in the structure, stability, dynamics and biological activity of the protein and has been a subject of enormous interest in the recent past.\cite{RN48,RN5,RN19,RN21,RN13,RN27,RN85,RN20,RN41,RN17,RN18,RN59,RN7,RN86,RN40,RN11,RN46,RN12,RN89} With the advent of new experimental\cite{RN62,RN38,RN37,RN71} and theoretical approaches\cite{RN27,RN89,RN4,RN56,RN2} many new aspects of this complex system have been unearthed. While the area of protein-water interactions have remained the focus of interest since the pioneering works of Pethig\cite{RN47}, Grant\cite{RN26}, W\"{u}thrich\cite{RN46,RN11} and others, a true quantification of hydration dynamics was achieved for the first time by the seminal work of Zewail and co-workers. In a series of pioneering studies, Professor Zewail confirmed the presence of an intermediate timescale component of time constant ~20-50 ps in the solvation dynamics of a natural probe.\cite{RN29,RN5,RN19,RN21,RN13,RN20,RN17,RN18} As a first, they used natural local probes (e.g. tryptophan) without disrupting the native states of proteins in several femtosecond resolved studies of solvation dynamics and reported a bimodal decay. Experimental studies on proteins like Subtilisin Carlsberg and Monellin typically show two primary relaxation times (one less than~1ps and the other in ~20-40 ps range).\cite{RN5,RN19} The slow component was attributed to the slow dynamics of biological water, term coined earlier by Nandi and Bagchi to articulate special properties of protein and DNA hydration layers. \cite{RN48} \par
Several  time dependent fluorescence Stokes shift\cite{RN5,RN19,RN21,RN20,RN2,RN58} (TDFSS) and three pulse photon echo peak shift\cite{RN62,RN64,RN65} (3PEPS) experiments with fluorescent dyes have revealed useful information about the dynamics in the immediate neighbourhood of the probe. TDFSS measures the instantaneous vibronic energy of the probe by perturbing the charge distribution using ultrafast lasers. From experimental data, a non-equilibrium Stokes shift response function S(t) is constructed \cite{RN67,RN16,RN71,RN56,RN37}(Eq.\ref{eq1})
\begin{equation} \label{eq1}
S(t) = \frac{{\nu (t) - \nu (\infty )}}{{\nu (0) - \nu (\infty )}} = \frac{{{E_{solv}}(t) - {E_{solv}}(\infty )}}{{{E_{solv}}(0) - {E_{solv}}(\infty )}}
\end{equation}
Here, $ \nu (t) $ is the time dependent fluorescence frequency of the probe at time $‘t’$,  proportional to $E_{solv}(t)$ which is the solvation energy of the probe (or solute) at time $‘t’$.
It is convenient to discuss the timescales involved in dipolar solvation dynamics. For an ion the solvation relaxation is much faster than dielectric relaxation. According to homogeneous dielectric continuum model, for an ion, the time constant of solvation $\tau_{L})$ is given by,
\begin{equation}\label{eq2}
{\tau _L} = \left( {\frac{{{\varepsilon_\infty }}}{{{\varepsilon _0}}}}\right){\tau _D}
\end{equation}
and that of a point dipole is given by \cite{RN83},
\begin{equation}
{\tau _L}^d= \left({\frac{{2{\varepsilon_\infty }+{\varepsilon_c}}}{{2{\varepsilon_0}+{\varepsilon_c}}}}\right){\tau _D}
\end{equation}
Where, $\tau_D$ is the Debye relaxation time, $\epsilon_c$ is the dielectric constant of the molecular cavity; $\epsilon_0$ and
$\epsilon_\infty$ are respectively the static and infinite frequency dielectric constants of the solvent. The dielectric constant 
not being well defined in the protein hydration layer, these expressions get modified. Solvochromatic studies often find that the 
dielectric constant of the hydration layer $\epsilon_{hyd}$, is less than that of the bulk, $\epsilon_{bulk}$. Hence, the process
of solvation slows down by a factor of two though there could be many other factors complicating the process. Castner \emph{et al.}
\cite{RN82} incorporated effects of inhomogeneity in the continuum model. However, 
predictions of the continuum models are found to be grossly inadequate for water. In reality the solvation dynamics of a probe in 
bulk water is extremely fast. In addition to an ultrafast component, there are two timescales $ \sim $40-50 fs and $\sim $1 ps.
\cite{RN46,RN12,RN83} The complex solvation in the bulk gets more complicated for hydration water. In addition to the multitude of 
timescales there is a coupled role of side-chain and water, which deserves proper quantification.\par
In the past few decades, several TDFSS measurements concentrating on probes bound to protein and DNA molecules have revealed a 
similar ultrafast component of amplitude of $ \sim $ 60\% . This has been attributed to librational and inter-molecular vibrational
modes. Besides there is a slow decay which is more pronounced than that in bulk water. However, the existence of this slow component 
is a much debated topic in this field. For example, earlier NMR studies by W\"{u}thrich \emph{et al}. have suggested a range of 
residence times of hydration water ($ \sim $300-500 ps to $ \sim $10-200 ps) which plays a major role in the dynamics of 
solvation.\cite{RN11,RN46} Later studies involving NMR and single particle orientation relaxation have contradicted the existence 
of slow component.\cite{RN32,RN42,RN22} However, dielectric relaxation[26,27,37] and solvation dynamics studies[21,23] which measure
collective responses show a considerable percentage of slow component.\par
Fleming \emph{et al.} used 3PEPS to examine solvation dynamics of eosin dye tied to protein surface. This revealed presence of two
distinct slow  timescales ($ \sim $100 ps and $ \sim $500 ps) that were not present in eosin-water system.\cite{RN62} Moreover, 
in a number of solvation dynamics studies, slow component within the range of $ \sim $100 - 1000 ps have been reported by 
Bhattacharyya \emph{et al.} \cite{RN27,RN85,RN86}\par
Frauenfelder \emph{et al.} have proposed ‘a unified model’ for protein dynamics from a series of Mössbauer spectroscopy 
experiments, which suggest slaving of small scale fluctuations in proteins by hydration layer fluctuations. Whereas, those in 
bulk water slave the large scale protein conformational fluctuations.\cite{RN49} Recently TDFSS experiments by Qin \emph{et al.} 
have found, using tryptophan as a probe, an ultrafast component around ~100 fs along with two slow components in the ~ps 
order.\cite{RN1}\par 
The simultaneous presence of the slow time scale in protein hydration dynamics and dielectric relaxation data seem to suggest 
a common origin. Earlier studies have attributed this component to the presence of a dynamic equilibrium between free and
quasi-bound water molecules in the hydration layer. One additional factor which often gets ignored is the role of charged 
amino acid side chains in the solvation process. Ali and Singer\cite{RN2} observed that if the motions of side-chains are 
quenched, relaxation becomes faster. Here we show that this apparently paradoxical result is actually a consequence of 
forced disappearance of a natural slow component. However, we find that the dependence on the motion of the amino acid 
side chain has no such universal characteristics. On quenching of the amino acid side chain motion, solvation can 
accelerate or decelerate depending on the nature and location of the probe.\par
Our present work focuses on the origin of slow relaxation in three model protein-water systems.  Our results are 
applicable to other biological macromolecules as well. \par
\section{THEORETICAL ANALYSIS}
Time dependent fluorescence frequency, $\nu(t)$ (in Eq.\ref{eq1}),  is a directly measurable quantity from TDFSS 
experiments. Despite $S(t)$ being a non-equilibrium response function, under the assumption that the solvent response
is linear to the external perturbation, we can equate $S(t)$ to the equilibrium energy autocorrelation function 
$C(t)$ (Eq.\ref{eq4}).\cite{RN37,RN69,RN70}
\begin{equation}\label{eq4}
C(t) = \frac{{ < \delta {E_{solv}}(0)\delta {E_{solv}}(t){ > _{gr}}}}{{ < \delta {E_{solv}}{{\left( 0 \right)}^2} > }}
\end{equation}
Here, $\delta E_{solv}(t)$ is the fluctuation of energy from its average value given by $\delta E_{solv}(t)=E_{solv}(t)- <E>$.
The subscript \emph{‘gr’} indicates averaging over ground state. Although there is no certainty that linear response theory 
would be valid for every system, generally under long time average ($\sim$50 ns or greater) the linear response correlation 
functions are in good agreement with experiments and simulations.\cite{RN2,RN4}\par
\begin{figure*}[htb!]
\begin{center}
\includegraphics[width=0.9\textwidth]{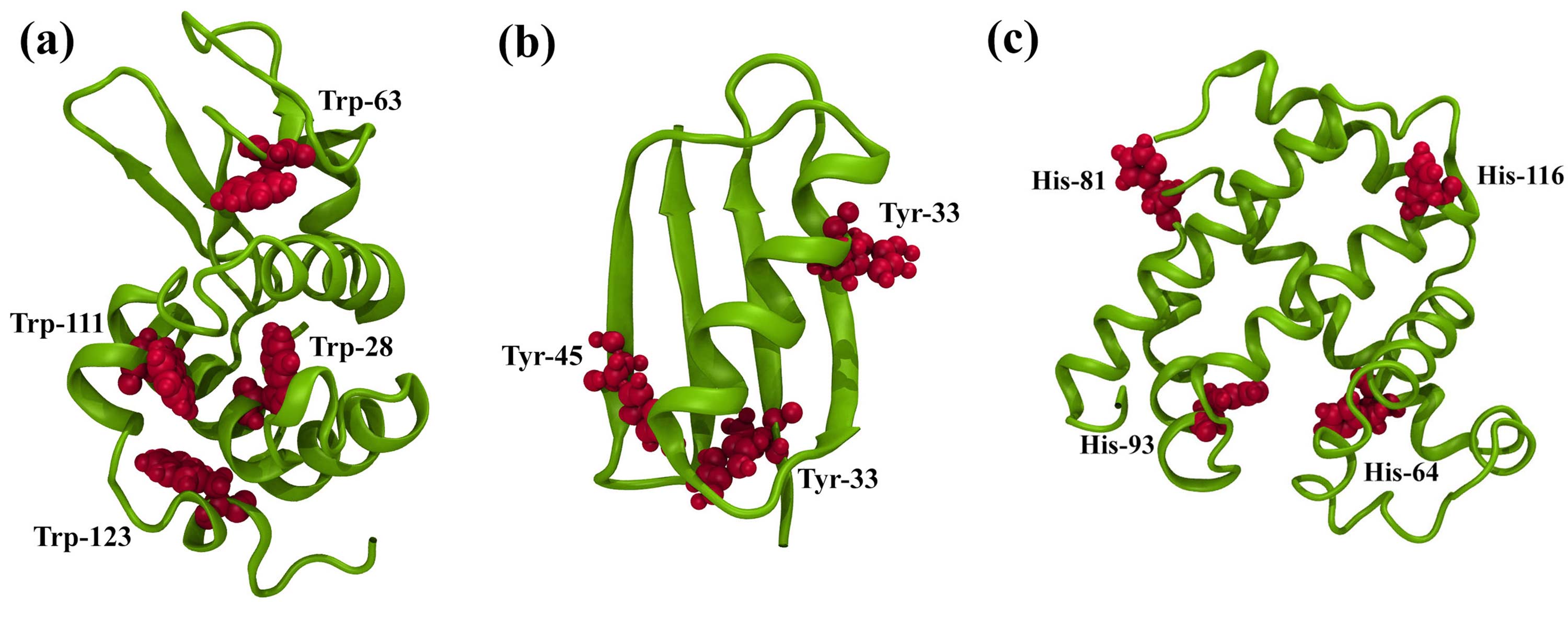}
\end{center}
\caption[]{\label{fig1}Ribbon representations of three proteins and locations of selected residues as natural probes. 
(a) Trp-28, Trp-63, Trp-111 and Trp-123 in Lysozyme (PDB: 1AKI), (b) Tyr-3, Tyr-33 and Tyr-45 in Protein G (PDB: GB1) 
(c) His-64, His-81, His-93 and His-116 in Sperm whale Myoglobin (PDB: 3E5O).}
\end{figure*}

 Unlike dielectric relaxation, solvation dynamics furnishes information on the local dynamics. With that spirit, we look 
 into various intrinsic probes in each protein to acquire information regarding the site-dependent timescales of relaxation. 
 The protein residues chosen in our study as intrinsic probes are shown in Fig.\ref{fig1}.\par
In addition to these natural probes, we also study the solvation energy relaxation of virtual spherical probes containi
ng one unit of positive charge, placed in different parts near the protein surface (inside the hydration layer) to mimic 
external fluorophores. The charges on different atoms of the proteins are acquired from OPLS-AA\cite{RN51} force field.
The time dependent solvation energy $E_{solv}(t)$ is decomposed into four parts, namely $E_{SC}(t)$, $E_{Core}(t)$, $E_{Wat}(t)$ 
and $E_{Ion}(t)$ to segregate the contributions of protein side-chains (SC), backbone (Core), solvent (Wat) and ions (Ion) to 
the total solvation response of a particular probe, as depicted in Eq. \ref{eq5} \cite{RN81,RN88}. Pal \emph{et al.} were the 
first to use this kind of decomposition method for a 38 base pair native DNA.\cite{RN81}
\begin{equation} \label{eq5}
E_{solv}(t)=E_{SC}(t)+E_{Core}(t)+E_{Wat}(t)+E_{Ion}(t)
\end{equation}
\begin{figure*}[htb!]
\begin{center}
\includegraphics[width=0.8\textwidth]{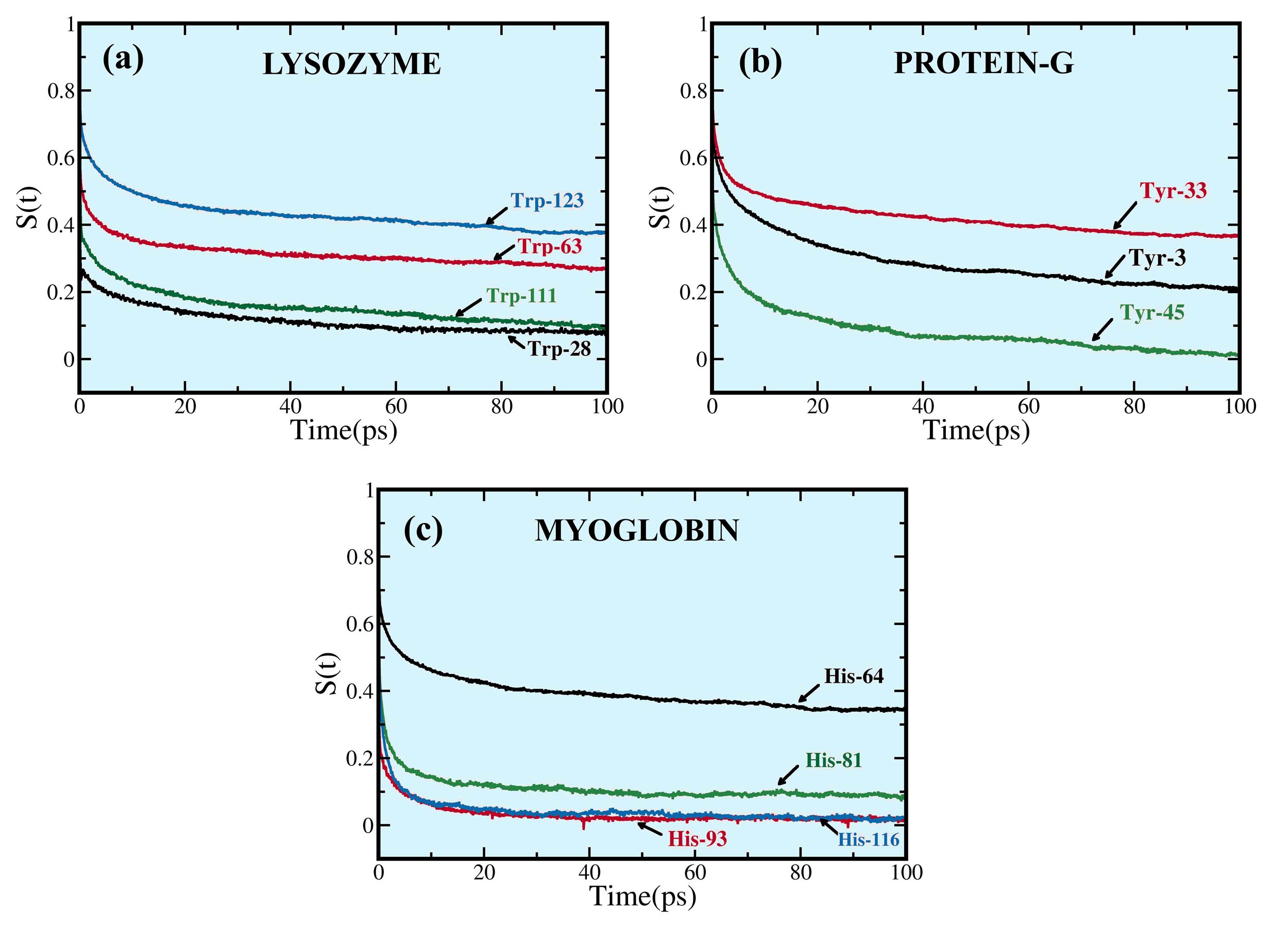}
\end{center}
\caption[]{\label{fig2}Normalised total solvation energy time correlation functions calculated from ground state 
equilibrium molecular dynamics simulations for intrinsic natural probes located in different parts of Lysozyme, 
Protein G and Myoglobin. (a)Trp-28, Trp-63, Trp-111 and Trp-123 in Lysozyme (b)Tyr-3, Tyr-33 and Tyr-45 in Protein-G (c)
His-64, His-81, His-93 and His-116 in Myoglobin.}
\end{figure*}
As there are a few number of counter ions (8 for lysozyme, 4 for protein-G and 2 for myoglobin) in our systems, their 
contribution to the total energy is negligible compared to others. Hence, the solvation time correlation function is a 
summation of three self and six cross-correlation terms (Eq.\ref{eq6}).
\begin{equation} \label{eq6}
S(t) = \sum\limits_\alpha{{S_{\alpha \alpha }}(t) + \sum\limits_\alpha  {\sum\limits_\beta{{S_{\alpha\beta}}}}}(t)
\end{equation}
Here, $\alpha$ and $\beta$ stand for different components, i.e., side-chain, core and water. We calculate $S(t)$ as well as $S_{\alpha \alpha}(t)$ and $S_{\alpha \beta}(t)$ from ground state equilibrium MD simulations so that their relative amplitudes can be compared to identify the dominant terms. The data are fitted to a multi-exponential function along with a gaussian component to find out the timescales of solvation energy relaxation (Eq.\ref{eq7}).\cite{RN71,RN4}
\begin{equation} \label{eq7}
S(t)={a_g}{e^{ - {{\left( {{\raise0.7ex\hbox{$t$} \!\mathord{\left/
 {\vphantom {t {{\tau _g}}}}\right.\kern-\nulldelimiterspace}
\!\lower0.7ex\hbox{${{\tau _g}}$}}} \right)}^2}}} + {\sum\limits_{i = 1}^n {{a_i}e} ^{ - \left( {{\raise0.7ex\hbox{$t$} \!\mathord{\left/
 {\vphantom {t {{\tau _i}}}}\right.\kern-\nulldelimiterspace}
\lower0.7ex\hbox{${{\tau _i}}$}}} \right)}}
\end{equation}
Average solvation time is calculated by integrating $S(t)$ with respect to time. As solvation dynamics is intimately connected to the orientation relaxation of the surrounding solvent, we calculate $r_{1}(t)$ and $r_{2}(t)$ (Eq.\ref{eq8} and \ref{eq9}) of the hydration layer and compare with bulk water.
\begin{equation} \label{eq8}
r_{1}(t)=<P_{1}(cos\theta(t))>
\end{equation}
\begin{equation} \label{eq9}
{r_2}(t)=\frac{2}{5} <{P_2}(cos\theta (t))>
\end{equation}
Here, $P_{1}$ and $P_{2}$ are respectively the first and second order Legendre polynomials. $\theta(t)$ is the angle between O\textemdash H 
bond vectors of water molecules at any arbitrary time $‘s’$ and the same at time $‘s+t’$. Further constrained MD simulations have 
been done by quenching the motions of protein atoms to isolate the effect of conformational fluctuations of protein on solvation 
energy relaxation.
\section{RESULTS AND DISCUSSION}
\subsection{Solvation dynamics correlation functions for different probes: Heterogeneity and multitude of dynamics}
In this work, we investigate the nature of solvation dynamics in each protein with respect to several side-chain residues selected
as intrinsic probes. The solvation energy correlation functions are averaged over three MD trajectories, each 50 ns long, starting 
from totally distinct conformations of the protein. Figure 2 shows the $S(t)$ plotted against time for the aforesaid probes.\par
It is clear from the results that the timescales of solvation are different for different probes even in the same protein molecule. 
Each of the probes have a sub $\sim$100 fs ultrafast component which is in good agreement with experiments\cite{RN2,RN1} and is 
arising out of the librational motion of water.\cite{RN27,RN4,RN5,RN85} In Lysozyme (Table \ref{table1} and Fig.\ref{fig2}a), the 
slow component is $\sim$400 ps for Trp-63 and Trp-123, almost three times higher than the slow component of Trp-111 and about twice 
that of Trp-28. Another intermediate timescale $\sim$10 ps for Trp-28 and $\sim$5 ps for other three is observed.\par
\begin{table*}[htbp!]
\begin{center}
\caption{\label{table1}Timeales of total solvation energy relaxation and respective SASA values for the four natural tryptophan
probes in Lysozyme-water system. Data are fitted using Eq.\ref{eq7}.}
\begin{ruledtabular}
\begin{tabular}{lllllllll}
%Probe & SASA ($nm^2$) & $a_g$ & $\tau_g$ (ps) 
Probe & SASA ($nm^2$) & $a_{g}$	& $\tau_g$ (ps)	& $a_1$	& $\tau_1$ (ps)	& $a_2$	& $\tau_2$ (ps) & $<\tau>$ (ps)\\ \hline 
Trp-28	& 0.06	& 0.75	& 0.08	& 0.12	& 11.9	& 0.13	& 188.1	& 25.9\\
Trp-63	& 0.36	& 0.50	& 0.08	& 0.16	& 5.4	& 0.34	& 446.6	& 152.7\\
Trp-111	& 0.32	& 0.62	& 0.09	& 0.17	& 5.7	& 0.21	& 132.8	& 28.9\\
Trp-123	& 0.69	& 0.32	& 0.09	& 0.20	& 5.44	& 0.47	& 414.9	& 196.1\\
\end{tabular}
\end{ruledtabular}
\end{center}
\end{table*} 
For Protein-G (Table \ref{table3}, supporting information (SI)), the differences in dynamics of the three tyrosine residues are also distinct. Tyr-33 has a slow component of 326.6 ps with 48\% contribution. In contrast, Tyr-45 has a slow component of around 44.1 ps with 19\% contribution whereas Tyr-3 falls in between Tyr-33 and Tyr-45 with a slow component of 167.1 ps (Fig.\ref{fig2}b).\par
Unlike Protein-G and Lysozyme, the dissimilarity is not that much pronounced in Myoglobin (Table \ref{table4}, SI). In spite of the different relative amplitudes of the partial components, the average relaxation times either remain almost same or differ only from one another by one order of magnitude. His-64 and His-81 exhibit slow timescales of 391.2 ps and 216.5 ps respectively. His-93 and His-116 have slow components of respectively 138.7 ps and 72.8 ps. Despite being located in different parts of the protein and having different solvent exposures, His-93 and His-116 shows almost similar relaxation behaviour (Fig.\ref{fig2}c). From the time averaged solvent accessible surface area (SASA) values\cite{RN74} of these residues no direct correlation can be drawn between the timescales of relaxation and the degree of solvent exposure. In some cases the relaxation of exposed residues is faster and in some others they are slower than the buried ones. It is observed that presence of polar/charged residues in immediate surroundings makes solvation dynamics noticeably slower.
\subsection{Sensitivity of solvation dynamics to the location of the probe}
	Analysis of the above data suggests that the solvation responses of different sites in the same protein can be different for the same probe. The local chemical and electrostatic environment of the probes along with the innate heterogeneity present in protein surfaces make solvation dynamics a region/domain dependent phenomenon. A closer look on the structure of the proteins reveals that the probes which are surrounded by polar-charged groups like arginine (Arg), lysine (Lys), aspartic acid (Asp), glutamic acid (Glu) etc. shows a significant slow component with arginine and lysine being the most effective ones. For instance, in lysozyme, Trp-28 which is buried deep inside the protein (SASA = 0.06 $nm^2$), having no charged atoms in its proximity, exhibits faster dynamics than Trp-123 which is considerably more exposed to the solvent (SASA = 0.69 $nm^2$) but at the same time surrounded by charged side chains such as Arg-5, Arg-114, Arg-125, Lys-33 and Asp-119. This pattern is also noticed for the other two systems. In myoglobin, His-64, which has one arginine (Arg-45) and two lysine residues (Lys-62, Lys-63) in its neighbourhood, shows slower dynamics compared to the other three probes. His-93 and His-116, being the most deprived with respect to charged groups in their vicinity, show a faster dynamics. These observations give a qualitative understanding of the relative slow dynamics of solvation.
\begin{figure*}[htbp!]
\begin{center}
\includegraphics[width=0.8\textwidth]{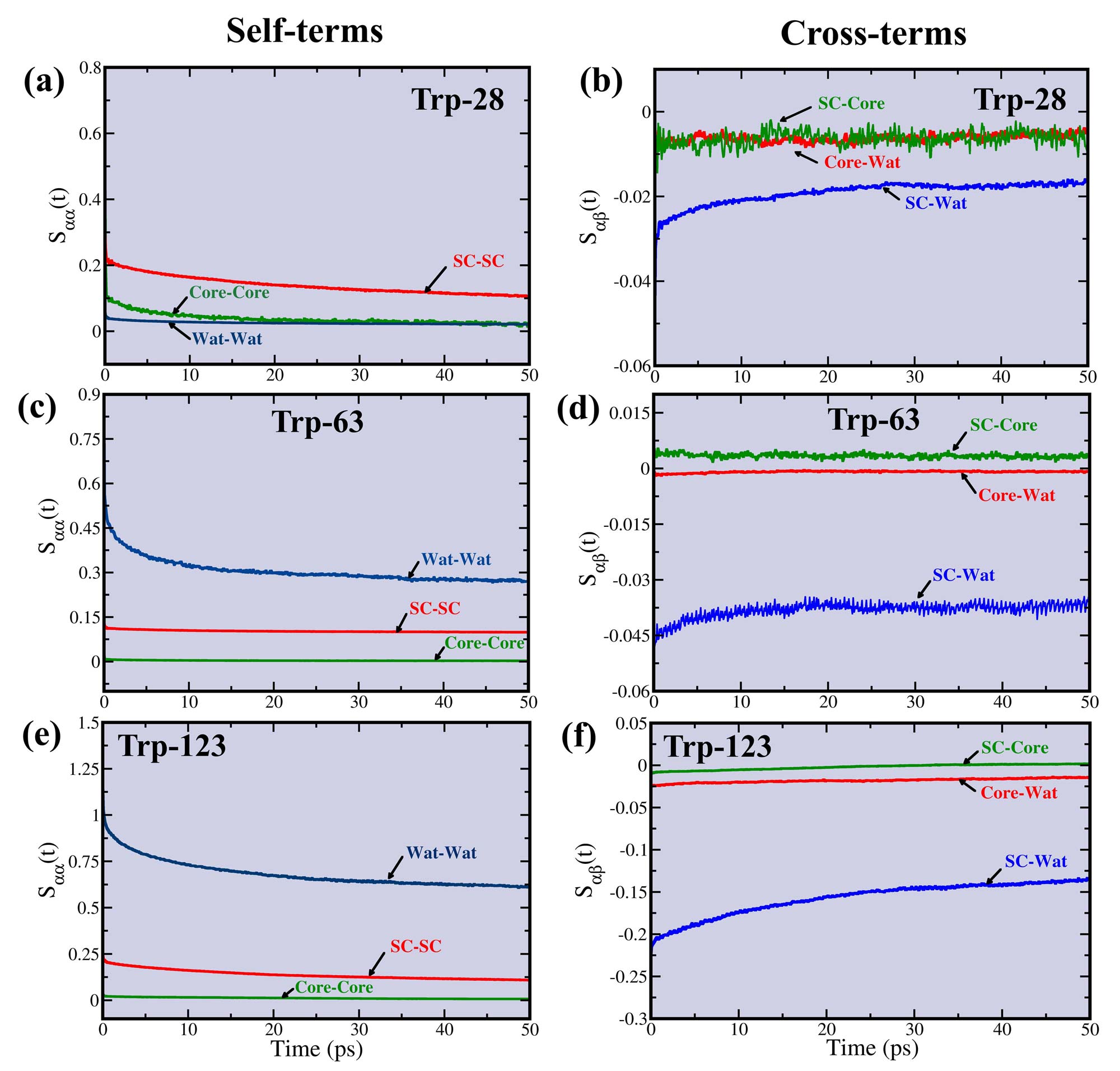}
\end{center}
\caption[]{\label{fig3}Plots of self and cross solvation energy correlation terms (scaled to the total solvation response) for three probes in Lysozyme which are calculated by decomposing the total solvation response into components of the system. $S_{\alpha\alpha}$(t) indicates the self-correlation terms where α can be either side-chain or water or protein-core. $S_{\alpha\beta}$(t) indicates the cross-correlation terms where $\alpha\beta$ is any combination of side-chain, water and protein-core $(\alpha\neq\beta)$. (a) Self terms for Trp-28, (b) Cross terms for Trp-28 (c) Self terms for Trp-63, (d) Cross terms for Trp-63, (e) Self terms for Trp-123 and (f) Cross terms for Trp-123.}
\end{figure*}
\begin{figure*}[htb!]
\begin{center}
\includegraphics[width=0.9\textwidth]{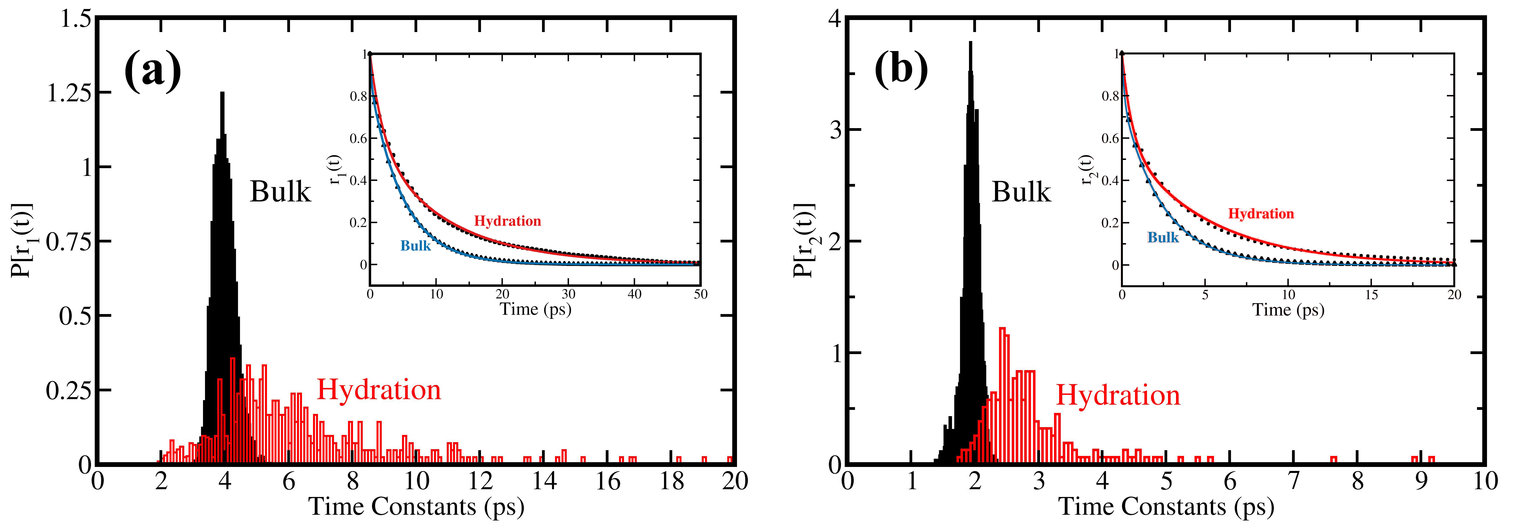}
\end{center}
\caption[]{\label{fig4}Histogram plots showing distribution of average orientation relaxation timescales for hydration and bulk water molecules. (a) Distribution of $r_1(t)$ and normalised time correlation function (inset) (b) Distribution of $r_2(t)$ and respective normalised time correlation function (inset). (The time-correlations were averaged over 500 hydration and 8000 bulk water molecules)}
\end{figure*}
\begin{figure*}[htbp!]
\begin{center}
\includegraphics[width=0.9\textwidth]{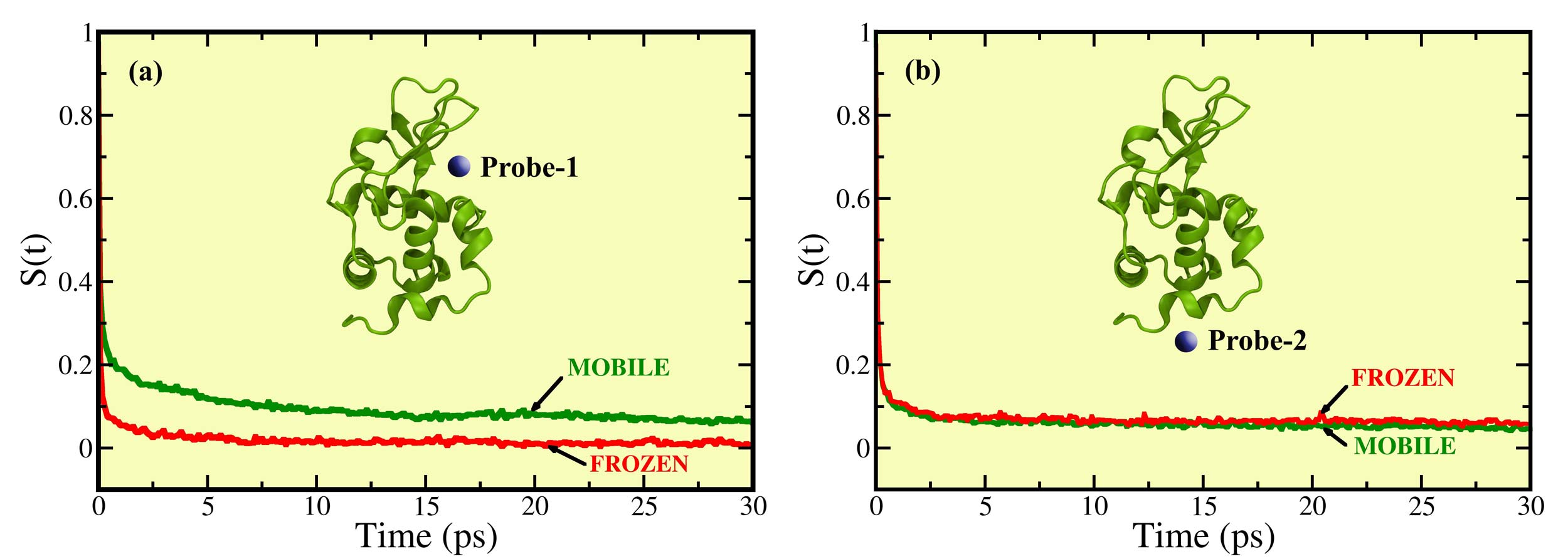}
\end{center}
\caption[]{\label{fig5}Results from constrained MD simulations which compares the total normalised solvation energy relaxation for two spherical virtual probes in lysozyme (a) Probe-1, placed near Trp-63 and Trp-62 i.e., the cavity region of Lysozyme shows faster relaxation upon freezing the protein motions (b) Probe-2, placed near Trp-123 of Lysozyme hardly shows any difference between the frozen and mobile protein cases.}
\end{figure*}
\subsection{Importance of the self and cross-correlation terms}
In order to explore the origin of dynamic heterogeneity in the hydration layer, we separate out the individual self and cross solvation energy correlation terms for these probes (Eq.\ref{eq6}). The results for three such representative probes (Trp-28, Trp-63 and Trp-123 for protein Lysozyme) are presented in Fig.\ref{fig3}.\par
It is noticeable that the cross terms have negative amplitudes which indicates anti-correlation and can be easily spotted in the energy trajectory. For example, the $S_{SC-Wat}(t)$ terms are always anti-correlated (Fig.\ref{fig3}), i.e., when the energy contribution coming from side-chains increases, that from water decreases and vice versa. This may originate because of side-chain assisted energy transfer from core to hydration layer. Screening effect may be responsible for this. When side-chain contribution to the solvation increases, the charges are not getting screened by the solvent. This in turn results in decreasing water contribution. On the other hand, when water molecules are free to orient rapidly, they contribute more to the solvation and contribution arising from side-chains decreases. One of the exposed probes, Trp-123 of Lysozyme derives most of its contribution from Wat-Wat and SC-SC self-terms. The origin of slow relaxation of the Wat-Wat and SC-Wat terms lies in the neighbourhood charges (section 3.2) which makes the water dynamics of that region slower by forming long lived hydrogen bonds with nearby water[46]. This effect is also reflected on the relative amplitudes of those terms. On the other hand, the core-core self-term has more dominant relative contribution than other terms in case of the buried probe Trp-28. As it is far away from the water environment the SC-Wat and Wat-Wat terms are fast decaying and of low relative amplitudes. A semi-exposed probe Trp-63 has slowly decaying Wat-Wat and SC-Wat terms which can be explained in similar fashion. For all the probes, the negative cross-correlations help the total solvation response decay at a faster rate. For Trp-28, the amplitudes of the cross-terms are almost negligible compared to self-terms. On the contrary, for Trp-123 the cross-terms play an important role in weakening the slow component but get overshadowed by the huge amplitude of Wat-Wat self-term.
\begin{figure*}[htbp!]
\begin{center}
\includegraphics[width=0.8\textwidth]{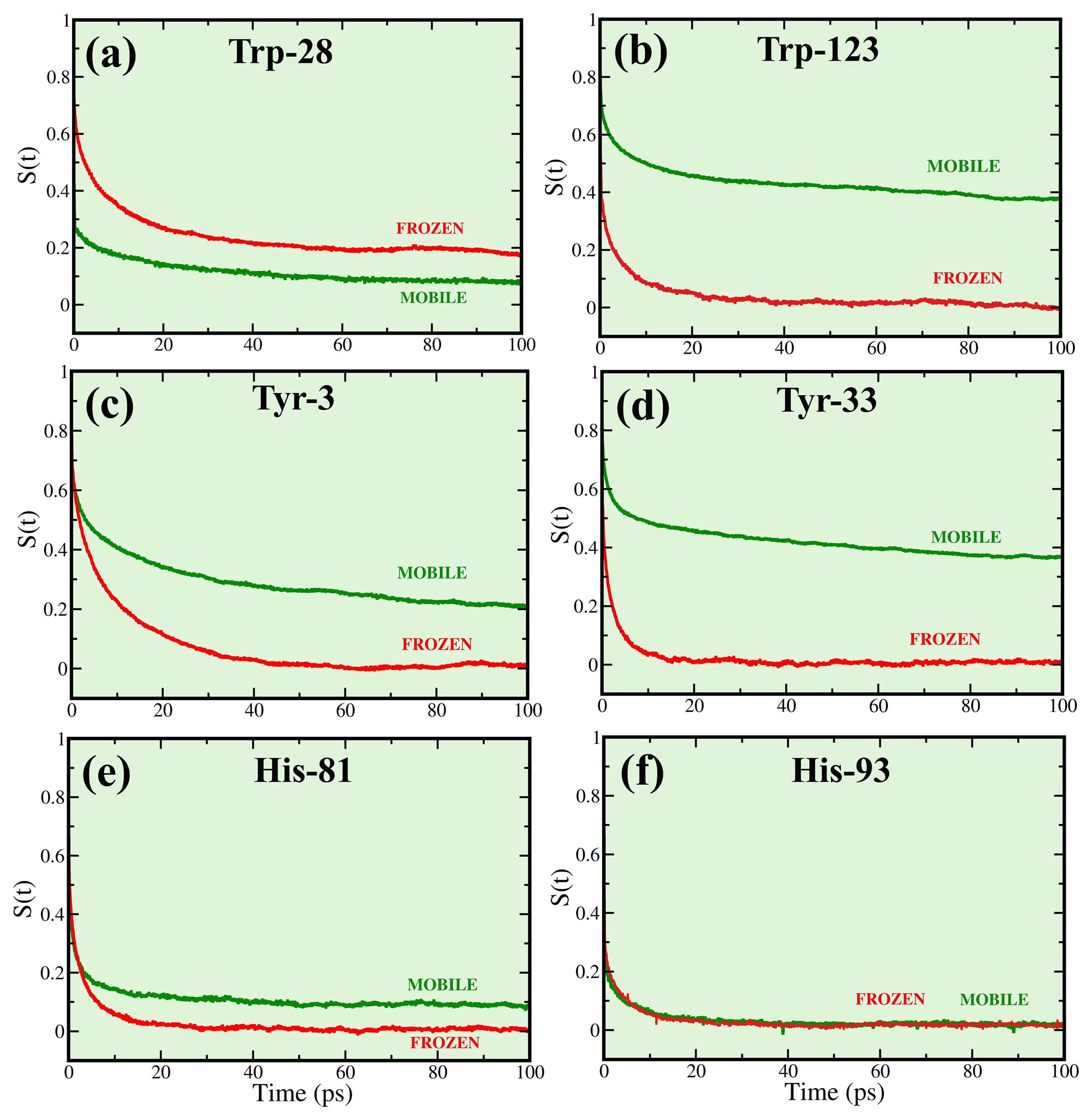}
\end{center}
\caption[]{\label{fig6}Comparison of solvation energy relaxation of natural probes between mobile and frozen protein cases. (a) Trp-28 of Lysozyme shows slower decay when protein motions are quenched. (b) Trp-123 of Lysozyme exhibits accelerated dynamics upon quenching the protein motions. (c) Tyr-3 of Protein-G and (d) Tyr-33 of Protein-G, both show faster dynamics when protein motions are absent. (e) His-81 of Myoglobin exhibits accelerated solvation upon freezing the protein. (f) His-93 of Myoglobin hardly shows any difference in solvation energy relaxation when protein is immobilised.}
\end{figure*}
\subsection{Orientational relaxation of water molecules in bulk and hydration layer}
The observed orientational relaxations of O\textemdash H bond vector, i.e. $r_{1}(t)$ and $r_{2}(t)$, are multi-exponential in nature. Average orientation relaxation time for hydration water molecules is almost twice that of the bulk water and the slow component for hydration water is almost ten times that of the bulk (Table \ref{table2}). From the histograms (Fig.\ref{fig4}a and\ref{fig4}b) it is clear that there exist rotationally faster water molecules inside the hydration layer (arising from weak hydrogen bonding and also from frequent random kicks from side-chain conformational fluctuations) than bulk along with slower ones. \emph{The existence of such disparate time scales was missed by NMR experiments as it measure average time scale and not a good technique to capture the various timescales of dynamics in protein hydration layer.} It seems that in constrained environment fast water molecules become faster and slow become slower. For faster solvation the water molecules have to orient/rotate at a faster rate. Sites exhibiting slow solvation are accompanied by slow orienting water molecules. We get both the signatures in the time correlation function (Table \ref{table2}) The multitude of timescales of orientational relaxation of the solvent is partly responsible for the observed heterogeneity in solvation dynamics.
\begin{table}[htbp!]
\begin{center}
\caption{\label{table2}Multi-exponential fitting parameters for orientation relaxation time correlation functions (as mentioned in Eq.\ref{eq8} and Eq.\ref{eq9}) for bulk and hydration water molecules in Lysozyme-water system.}
\begin{ruledtabular}
\begin{tabular}{llllll}
$r_{1}$(t)	& $a_1$	& $\tau_1$ (ps)	& $a_2$	& $\tau_2$ (ps)	& $<\tau>$ (ps)\\ \hline
Hydration	& 0.58	& 11.47	& 0.42	& 2.01	& 7.49\\
Bulk	& 0.87	& 4.93	& 0.13	& 0.21	& 4.32\\ \hline
$r_{2}$(t)	& $a_1$	& $\tau_1$ (ps)	& $a_2$	& $\tau_2$ (ps)	& $<\tau>$ (ps)\\ \hline
Hydration	& 0.57	& 5.07	& 0.43	& 0.59	& 3.14\\
Bulk	& 0.78	& 2.39	& 0.22	& 0.13	& 1.89\\
\end{tabular}
\end{ruledtabular}
\end{center}
\end{table} 
\subsection{Role of the side-chain conformational fluctuations in solvation dynamics}
To understand the dependence of solvation energy relaxation on the conformational fluctuation of side chains, we simulate the same systems by artificially freezing the coordinates of the protein atoms so that the protein resides in the system as a rigid charge distribution but still interacting with its surroundings. It is seen that the solvation of the virtual spherical probe located in the cavity region of lysozyme (Probe-1) near Trp-63 gets accelerated upon freezing the protein motions but no such significant change is observed for the other one (Probe-2) located near Trp-123 (Table \ref{table5}, SI). Conformational fluctuation of protein contributes to the slow solvation near the vicinity of Probe-1 (Fig. \ref{fig5}a). On the contrary, the solvation energy relaxation does not depend on the protein motion in the locality of Probe-2 (Fig. \ref{fig5}b). Thus we can distinguish two entirely different domains in the hydration layer of Lysozyme with respect to the observed contrasting dynamics.\par
Dynamics of the natural probes show similar heterogeneity. When the protein motions are quenched the total solvation becomes faster in case of His-81 (Fig.\ref{fig6}e) and Trp-123 (Fig.\ref{fig6}b). But no such noticeable effect is observed for His-93 (Fig.\ref{fig6}f). However an inverse effect is seen in case of Trp-28 of Lysozyme (Fig.\ref{fig6}a). That is the relaxation is slower for the frozen protein case. However, Tyr-3 (Fig.\ref{fig6}c) and Tyr-33 (Fig.\ref{fig6}d) of protein-G exhibits slower decay which is dependent on side-chain motion. This clearly indicates that the slow component in solvation dynamics arises partly from protein motions, particularly from side-chain conformational fluctuations involving charged groups. This also supports the observation of Singer \emph{et al.}\cite{RN2} Other probes in all three proteins show, though not to the same extent, accelerated dynamics upon freezing the protein. However, It is noticeable, though the solvation becomes faster, the intermediate timescales, as observed by Zewail \emph{et al.}, does not vanish completely (Table \ref{table6}, SI).\par
The role of water molecules is somewhat difficult to separate out as they are coupled to the motion of the side chain atoms themselves.  We thus surmise that whenever a probe is surrounded by polar-charged residues, the nearby water molecules participate in hydrogen bonding with the residues present in the neighbourhood. From hydrogen bond dynamics studies it becomes clear that the hydrogen bond lifetime is the highest for polar-charged residues like arginine and lysine.[46] So, the presence of such residues in the vicinity of a particular probe make the local water dynamics slower and as a result, slow solvation is observed.  When it comes to solvation dynamics, location and neighbourhood of the probe is also a governing factor along with the intrinsic nature and degree of solvent exposure of the probe.\par 
 It is thus somewhat paradoxical that the conformational fluctuations of amino acid side chains, especially the charged ones, seem to make the solvation dynamics slower. However, this is not universally true from the results that we have obtained. (Table \ref{table5}, SI) This distinct behaviour of relaxation and dependence of the slow component on the side chain motions can again be attributed to the heterogeneity of the protein surface and hydration layer leading to the complex nature of dynamics across the different regions of a protein.
 \newpage
\section{CONCLUSIONS}
In a series of influential papers, Ahmed Zewail and co-workers employed natural probes that allowed unambiguous and systematic study of the unperturbed dynamics of protein hydration layer. They firmly established the presence of an intermediate range time scale in protein hydration dynamics. The  existence of  such a time scale was a subject of considerable controversy as the NMR experiments of Halle and co-workers\cite{RN22,RN23,RN24,RN25,RN32} and also several simulation studies have suggested the absence of such a time scale. Zewail established beyond doubt that such a time scale indeed exists in the solvation dynamics of protein hydration later. Bhattacharyya and co-workers found slower decays which are also getting verified lately \cite{RN27,RN85,RN86}. However, the origin of the discrepancy was not clear. Hopefully, our present study helps in explaining the origin of the reported differences.\par
What could possibly be the origin of such different results that are obtained in NMR-based studies and also in simulations vis-\`{a}-vis solvation dynamics and dielectric relaxation? There are several factors that could be responsible; (i) NMR studies cannot distinguish multiple disparate time scales and always give an average time scale. As we pointed out elsewhere that the presence of a 20\% of relaxation time that is slower by a factor of 10 or more than the usual bulk water relaxation can make the average relaxation slower by only a factor of 4-5. This explains much of the NMR disagreement with Zewail’s data. (ii) Hydration layer differs greatly from protein to protein. Besides, the fraction of slow water is often small. As from our studies we found ~15\% transnationally and ~27\% rotationally slow water in the hydration layer of Lysozyme (Fig. \ref{fig4}a and \ref{fig4}b). Moreover, a fraction of water molecules in the layer exhibits faster than bulk dynamics. So, the average becomes a poor measure of the time scale of dynamics. (iii) NMR and simulations studies focus on single particle dynamics while dielectric relaxation and solvation dynamics measure collective properties. (iv) Solvation dynamics probes energy relaxation that derives a significant contribution from polar side-chain motions which is discussed below in a bit more detail.\par
Multiple time scales observed here are seen to originate from various sources. We establish that the probes surrounded by charged residues exhibit slower dynamics which is because of the long lived hydrogen bonds resulting in quasi bound water molecules in that region making the orientation of water molecules restricted. The decomposition of total solvation energy yields several self and cross-correlations out of which the cross-terms are anti-correlated. The slowness in the ‘Wat-Wat’ and ‘SC-Wat’ terms arise for residues accompanied by other charged groups which participate in hydrogen bonding.\par
Our study also reveals the role of conformational fluctuations in solvation. In some parts (rich in charged atoms) the side-chain motions make the solvation faster whereas in some other parts (buried, deprived of charges in the vicinity) it decelerates or doesn’t affect the same. The relatively slow solvation by hydrogen bonded water to the polar/charged protein-atoms follow the dynamical exchange model.\cite{RN48} When the contribution of polar side-chains becomes important, the contribution of slow (bound) water also increases. When these motions are quenched, the slowest time scale reduces drastically, although solvation dynamics remains substantially slower than that in bulk water. The remaining slowness is arising from water molecules which are still hydrogen bonded to polar/charged atoms of the side chain. It is only possible to partly separate them. However, to what extent these contributions are coupled with side-chain fluctuations, still attracts further investigations. Some rigorous works are in progress to determine the molecular origin of slow dynamics.[46] Yet, the slow contribution doesn’t always come from the side-chains. There are further scope of detailed theoretical investigation and formulation of new parameters to untangle the complexity of the problem.
\section{SIMULATION METHODS}
Atomistic molecular dynamics simulations have been performed using GROMACS package\cite{RN87}. We have constructed the systems to match the experimental concentration ($\sim$2-3 mM). The initial configurations of the proteins have been taken from crystal structures available in Protein Data Bank. We have used OPLS-AA force field\cite{RN51} and extended point charge (SPC/E) water model. Periodic boundary conditions were implemented using cubic boxes of sides 94\AA  with 26,338 water molecules for Lysozyme (PDB ID: 1AKI); 90 \AA with 23,960 water molecules for immunoglobin binding protein G (PDB ID: GB1) and 93 \AA  with 26234 water molecules for sperm whale myoglobin (PDB ID: 3E5O).\par
The total system was energy minimised using steepest descend followed by conjugate gradient method. The solvent was equilibrated for 10 ns at constant temperature (300 K) and pressure (1 bar) (NPT) by restraining the positions of the protein atoms followed by equilibrium without position restrain for another 10 ns. The final production runs were carried out at a constant temperature (T=300 K) (NVT) for 55 ns. For analysis, the trajectories were recorded for the last 50 ns with 10 fs resolution. he equations of motions were integrated using leap-frog integrator with an MD time step of 0.5 fs.\par
We have used the N\'{o}se-Hoover\cite{RN53,RN54,RN55} thermostat and Parrinello-Rahman barostat\cite{RN78} to keep the temperature and pressure constant respectively. The cut-off radius for neighbour searching and non-bonded interactions was taken to be 10\AA and all the bonds were constrained using the LINCS algorithm\cite{RN76}. For the calculation of electrostatic interactions, Particle Mesh Ewald (PME)\cite{RN77} was used with FFT grid spacing of 1.6\AA.
\section{ACKNOWLEDGEMENTS}
We are grateful to Professor Ahmed H Zewail for his leadership,collaboration and contribution in the area of chemical dynamics. We thank Dr. Rajib Biswas for many fruitful discussions.  We gratefully acknowledge the support by grants from DST and CSIR, India. B. Bagchi acknowledges sir J C Bose fellowship. S. Mondal and S. Mukherjee also acknowledge the financial support obtained from UGC and INSPIRE, India.\par

\section{SUPPORTING INFORMATION}
In this section the multiexponential fitting data for various natural probes in two proteins (Protein-G and Myoglobin) which are discussed in the main text are provided (Table \ref{table3} and \ref{table4}). Also the timescales for two virtual probes inside the hydration layer of Lysozyme (both, when mobile as well as frozen state) are noted down (Table \ref{table5}). At the end, data for comparative study; i.e, between mobile and frozen proten states; of solvation on six representative natural probes in three proteins are given.(Table \ref{table6})  
\begin{table*}[htbp!]
\begin{center}
\caption{\label{table3}: Timescales of total solvation energy relaxation, S(t), of the three tyrosine residues in Protein G along with the respective SASA values.}
\begin{ruledtabular}
\begin{tabular}{llllllllll}
Protein & Probe & SASA ($nm^2$) & $a_{g}$	& $\tau_g$ (ps)	& $a_1$	& $\tau_1$ (ps)	& $a_2$	& $\tau_2$ (ps) & $<\tau>$ (ps)\\ \hline 
Protein G	& Tyr-3	& 0.13	& 0.39	& 0.08	& 0.24	& 6.7	& 0.37	& 167.1	& 63.5\\
 & Tyr-33	& 0.98	& 0.32	& 0.09	& 0.20	& 3.5	& 0.48	& 326.6	& 157.5\\
 & Tyr-45	& 0.89	& 0.54	& 0.07	& 0.27	& 3.3	& 0.19	& 44.1	& 9.3\\
\end{tabular}
\end{ruledtabular}
\end{center}
\end{table*}
 
\begin{table*}[htbp!]
\begin{center}
\caption{\label{table4}: Timescales of total solvation energy relaxation, S(t), of the three histidine residues in Myoglobin along with the respective SASA values.}
\begin{ruledtabular}
\begin{tabular}{llllllllll}
Protein & Probe & SASA ($nm^2$) & $a_{g}$	& $\tau_g$ (ps)	& $a_1$	& $\tau_1$ (ps)	& $a_2$	& $\tau_2$ (ps) & $<\tau>$ (ps)\\ \hline 
Myoglobin	& His-64	& 0.01	& 0.37	& 0.09	& 0.20	& 5.9	& 0.43	& 391.2	& 169.4\\
& His-81	& 1.14	& 0.59	& 0.09	& 0.28	& 2.7	& 0.13	& 216.5	& 28.9\\
& His-93	& 0.25	& 0.79	& 0.07	& 0.18	& 5.2	& 0.03	& 138.7	& 5.1\\
& His-116	& 0.86	& 0.60	& 0.08	& 0.34	& 2.1	& 0.06	& 72.8	& 5.1\\
\end{tabular}
\end{ruledtabular}
\end{center}
\end{table*}

\begin{table*}[htbp!]
\begin{center}
\caption{\label{table5}: Timescales of S(t) for two virtual spherical probes placed in two different regions near the protein surface.}
\begin{ruledtabular}
\begin{tabular}{llllllllll}
Probes & Protein & State of protein & $a_{g}$	& $\tau_g$ (ps)	& $a_1$	& $\tau_1$ (ps)	& $a_2$	& $\tau_2$ (ps) & $<\tau>$ (ps)\\ \hline 
Probe-1 &       Lysozyme &      Mobile &        0.77 &  0.08 &  0.15 &  4.02 &  0.08 &  137.6 & 11.6\\
        &                &      Frozen &        0.91 &  0.07 &  0.08 &  1.94 &  0.01 &  184.8 & 2.1\\
Probe-2 &       Lysozyme &      Mobile &        0.82 &  0.07 &  0.12 &  1.21 &  0.06 &  154.6 & 9.5\\
        &                &      Frozen &        0.82 &  0.08 &  0.11 &  1.14 &  0.07 &  419.5 & 29.5\\
\end{tabular}
\end{ruledtabular}
\end{center}
\end{table*}

\begin{table*}[htbp!]
\begin{center}
\caption{\label{table6}: Timescales of S(t) noted down for six intrinsic probes, two in lysozyme (Trp-28 and Trp-63), two in Protein-G (Tyr-3 and Tyr-33) and two in myoglobin (His-81 and His-93). A comparison of relaxation patterns between mobile and frozen protein cases.}
\begin{ruledtabular}
\begin{tabular}{llllllllll}
Probe    &      Protein         &  State of Protein  &          $a_g$      &   $\tau_g$ (ps)      & $a_1$     & $\tau_1$ (ps)     & $a_2$    & $\tau_2$ (ps) & $<\tau>$ (ps)\\ \hline
Trp-28   &      Lysozyme        &  Mobile            &          0.75    &   0.08         & 0.12   & 11.9        & 0.13  & 188.1 & 25.9\\
         &                      &  Frozen            &          0.40    &   0.09         & 0.36   & 8.7         & 0.24  & 343.2 & 85.5\\
Trp-123  &      Lysozyme        &  Mobile            &          0.32    &   0.09         & 0.20   & 5.4         & 0.47  & 414.9 & 196.1\\
         &                      &  Frozen            &          0.82    &   0.09         & 0.16   & 3.1         & 0.02  & 18.7  & 0.9\\
Tyr-3    &      Protein-G       &  Mobile            &          0.39    &   0.08         & 0.24   & 6.7         & 0.37  & 167.1 & 63.5\\
         &                      &  Frozen            &          0.31    &   0.08         & 0.24   & 2.1         & 0.45  & 14.4  & 7.0\\
Tyr-33   &      Protein-G       &  Mobile            &          0.32    &   0.09         & 0.20   & 3.5         & 0.48  & 326.6 & 157.5\\
         &                      &  Frozen            &          0.53    &   0.08         & 0.44   & 2.6         & 0.03  & 42.6  & 2.5\\
His-81   &      Myoglobin       &  Mobile            &          0.59    &   0.09         & 0.28   & 2.7         & 0.13  & 216.5 & 28.9\\
         &                      &  Frozen            &          0.51    &   0.09         & 0.42   & 2.7         & 0.07  & 21.6  & 2.7\\
His-93   &      Myoglobin       &  Mobile            &          0.79    &   0.07         & 0.18   & 5.2         & 0.03  & 138.7 & 5.1\\
         &                      &  Frozen            &          0.73    &   0.08         & 0.25   & 4.6         & 0.02  & 197.6 & 5.1\\
\end{tabular}
\end{ruledtabular}
\end{center}
\end{table*}

\newpage
\bibliography{SD_library}

\begin{thebibliography}{10}

\bibitem{RN48}
N.~Nandi and B.~Bagchi, ``Dielectric relaxation of biological water,'' {\em J.
  Phys. Chem. B}, vol.~101, no.~50, pp.~10954--10961, 1997.

\bibitem{RN5}
S.~K. Pal, J.~Peon, B.~Bagchi, and A.~H. Zewail, ``Biological water:
  femtosecond dynamics of macromolecular hydration,'' {\em J. Phys. Chem. B},
  vol.~106, no.~48, pp.~12376--12395, 2002.

\bibitem{RN19}
S.~K. Pal, J.~Peon, and A.~H. Zewail, ``Biological water at the protein
  surface: dynamical solvation probed directly with femtosecond resolution,''
  {\em Proc. Natl. Acad. Sci. U.S.A}, vol.~99, no.~4, pp.~1763--1768, 2002.

\bibitem{RN21}
S.~K. Pal, J.~Peon, and A.~H. Zewail, ``Ultrafast surface hydration dynamics
  and expression of protein functionality: α-chymotrypsin,'' {\em Proc. Natl.
  Acad. Sci. U.S.A}, vol.~99, no.~24, pp.~15297--15302, 2002.

\bibitem{RN13}
S.~K. Pal and A.~H. Zewail, ``Dynamics of water in biological recognition,''
  {\em Chem. Rev.}, vol.~104, no.~4, pp.~2099--2124, 2004.

\bibitem{RN27}
K.~Bhattacharyya, ``Solvation dynamics and proton transfer in supramolecular
  assemblies,'' {\em Acc. Chem. Res.}, vol.~36, no.~2, pp.~95--101, 2003.

\bibitem{RN85}
K.~Bhattacharyya, ``Nature of biological water: a femtosecond study,'' {\em
  Chem. Comm.}, no.~25, pp.~2848--2857, 2008.

\bibitem{RN20}
J.~Peon, S.~K. Pal, and A.~H. Zewail, ``Hydration at the surface of the protein
  monellin: dynamics with femtosecond resolution,'' {\em Proc. Natl. Acad. Sci.
  U.S.A}, vol.~99, no.~17, pp.~10964--10969, 2002.

\bibitem{RN41}
S.~M. Bhattacharyya, Z.-G. Wang, and A.~H. Zewail, ``Dynamics of water near a
  protein surface,'' {\em J. Phys. Chem. B}, vol.~107, no.~47,
  pp.~13218--13228, 2003.

\bibitem{RN17}
D.~Zhong, S.~K. Pal, and A.~H. Zewail, ``Biological water: A critique,'' {\em
  Chem. Phys. Lett.}, vol.~503, no.~1, pp.~1--11, 2011.

\bibitem{RN18}
D.~Zhong, S.~K. Pal, D.~Zhang, S.~I. Chan, and A.~H. Zewail, ``Femtosecond
  dynamics of rubredoxin: Tryptophan solvation and resonance energy transfer in
  the protein,'' {\em Proc. Natl. Acad. Sci. U.S.A}, vol.~99, no.~1,
  pp.~13--18, 2002.

\bibitem{RN59}
W.~Qiu, L.~Zhang, O.~Okobiah, Y.~Yang, L.~Wang, D.~Zhong, and A.~H. Zewail,
  ``Ultrafast solvation dynamics of human serum albumin: correlations with
  conformational transitions and site-selected recognition,'' {\em J. Phys.
  Chem. B}, vol.~110, no.~21, pp.~10540--10549, 2006.

\bibitem{RN7}
B.~Bagchi, ``Untangling complex dynamics of biological water at protein–water
  interface,'' {\em Proc. Natl. Acad. Sci. U.S.A}, vol.~113, no.~30,
  pp.~8355--8357, 2016.

\bibitem{RN86}
K.~Bhattacharyya and B.~Bagchi, ``Slow dynamics of constrained water in complex
  geometries,'' {\em J. Phys. Chem. A}, vol.~104, no.~46, pp.~10603--10613,
  2000.

\bibitem{RN40}
B.~Bagchi, {\em Water in Biological and Chemical Processes: From Structure and
  Dynamics to Function}.
\newblock Cambridge University Press, 2013.

\bibitem{RN11}
G.~Otting, E.~Liepinsh, and K.~W\"{u}thrich, ``Protein hydration in aqueous
  solution,'' {\em Science}, vol.~254, no.~5034, pp.~974--980, 1991.

\bibitem{RN46}
K.~W\"{u}thrich, M.~Billeter, P.~Güntert, P.~Luginbühl, R.~Riek, and
  G.~Wider, ``Nmr studies of the hydration of biological macromolecules,'' {\em
  Faraday Discuss.}, vol.~103, pp.~245--253, 1996.

\bibitem{RN12}
Y.~Levy and J.~N. Onuchic, ``Water mediation in protein folding and molecular
  recognition,'' {\em Annu. Rev. Biophys. Biomol. Struct.}, vol.~35,
  pp.~389--415, 2006.

\bibitem{RN89}
S.~Bandyopadhyay, S.~Chakraborty, and B.~Bagchi, ``Secondary structure
  sensitivity of hydrogen bond lifetime dynamics in the protein hydration
  layer,'' {\em J. Amer. Chem. Soc}, vol.~127, no.~47, pp.~16660--16667, 2005.

\bibitem{RN62}
X.~J. Jordanides, M.~J. Lang, X.~Song, and G.~R. Fleming, ``Solvation dynamics
  in protein environments studied by photon echo spectroscopy,'' {\em J. Phys.
  Chem. B}, vol.~103, no.~37, pp.~7995--8005, 1999.

\bibitem{RN38}
M.~Maroncelli and G.~R. Fleming, ``Picosecond solvation dynamics of coumarin
  153: the importance of molecular aspects of solvation,'' {\em J. Chem.
  Phys.}, vol.~86, no.~11, pp.~6221--6239, 1987.

\bibitem{RN37}
M.~Maroncelli and G.~R. Fleming, ``Computer simulation of the dynamics of
  aqueous solvation,'' {\em J. Chem. Phys.}, vol.~89, no.~8, pp.~5044--5069,
  1988.

\bibitem{RN71}
R.~Jimenez, G.~R. Fleming, P.~Kumar, and M.~Maroncelli, ``dynamics of water,''
  {\em Nature}, vol.~369, pp.~471--473, 1994.

\bibitem{RN4}
B.~Bagchi and B.~Jana, ``Solvation dynamics in dipolar liquids,'' {\em Chem.
  Soc. Rev.}, vol.~39, no.~6, pp.~1936--1954, 2010.

\bibitem{RN56}
M.~Maroncelli, ``Computer simulations of solvation dynamics in acetonitrile,''
  {\em J. Chem. Phys.}, vol.~94, no.~3, pp.~2084--2103, 1991.

\bibitem{RN2}
T.~Li, A.~A. Hassanali, Y.-T. Kao, D.~Zhong, and S.~J. Singer, ``Hydration
  dynamics and time scales of coupled water-protein fluctuations,'' {\em J.
  Amer. Chem. Soc}, vol.~129, no.~11, pp.~3376--3382, 2007.

\bibitem{RN47}
R.~Pethig, ``Protein-water interactions determined by dielectric methods,''
  {\em Annu. Rev. Phys. Chem.}, vol.~43, no.~1, pp.~177--205, 1992.

\bibitem{RN26}
E.~Grant, ``The dielectric method of investigating bound water in biological
  material: An appraisal of the technique,'' {\em Bioelectromagnetics}, vol.~3,
  no.~1, pp.~17--24, 1982.

\bibitem{RN29}
S.~K. Pal, D.~Mandal, D.~Sukul, S.~Sen, and K.~Bhattacharyya, ``Solvation
  dynamics of dcm in human serum albumin,'' {\em J. Phys. Chem. B}, vol.~105,
  no.~7, pp.~1438--1441, 2001.

\bibitem{RN58}
P.~Abbyad, X.~Shi, W.~Childs, T.~B. McAnaney, B.~E. Cohen, and S.~G. Boxer,
  ``Measurement of solvation responses at multiple sites in a globular
  protein,'' {\em J. Phys. Chem. B}, vol.~111, no.~28, pp.~8269--8276, 2007.

\bibitem{RN64}
W.~P. de~Boeij, M.~S. Pshenichnikov, and D.~A. Wiersma, ``Ultrafast solvation
  dynamics explored by femtosecond photon echo spectroscopies,'' {\em Annu.
  Rev. Phys. Chem.}, vol.~49, no.~1, pp.~99--123, 1998.

\bibitem{RN65}
J.~T. Kennis, D.~S. Larsen, K.~Ohta, M.~T. Facciotti, R.~M. Glaeser, and G.~R.
  Fleming, ``Ultrafast protein dynamics of bacteriorhodopsin probed by photon
  echo and transient absorption spectroscopy,'' {\em J. Phys. Chem. B},
  vol.~106, no.~23, pp.~6067--6080, 2002.

\bibitem{RN67}
B.~Bagchi, ``Dynamics of solvation and charge transfer reactions in dipolar
  liquids,'' {\em Annu. Rev. Phys. Chem.}, vol.~40, no.~1, pp.~115--141, 1989.

\bibitem{RN16}
B.~Bagchi, ``Water dynamics in the hydration layer around proteins and
  micelles,'' {\em Chem. Rev.}, vol.~105, no.~9, pp.~3197--3219, 2005.

\bibitem{RN83}
B.~Bagchi, D.~W. Oxtoby, and G.~R. Fleming, ``Theory of the time development of
  the stokes shift in polar media,'' {\em Chem. Phys.}, vol.~86, no.~3,
  pp.~257--267, 1984.

\bibitem{RN82}
E.~W. Castner~Jr, G.~R. Fleming, B.~Bagchi, and M.~Maroncelli, ``The dynamics
  of polar solvation: inhomogeneous dielectric continuum models,'' {\em J.
  Chem. Phys.}, vol.~89, no.~6, pp.~3519--3534, 1988.

\bibitem{RN32}
B.~Halle and L.~Nilsson, ``Does the dynamic stokes shift report on slow protein
  hydration dynamics?,'' {\em J. Phys. Chem. B}, vol.~113, no.~24,
  pp.~8210--8213, 2009.

\bibitem{RN42}
D.~Laage, G.~Stirnemann, and J.~T. Hynes, ``Why water reorientation slows
  without iceberg formation around hydrophobic solutes,'' {\em J. Phys. Chem.
  B}, vol.~113, no.~8, pp.~2428--2435, 2009.

\bibitem{RN22}
L.~Nilsson and B.~Halle, ``Molecular origin of time-dependent fluorescence
  shifts in proteins,'' {\em Proc. Natl. Acad. Sci. U.S.A}, vol.~102, no.~39,
  pp.~13867--13872, 2005.

\bibitem{RN49}
P.~W. Fenimore, H.~Frauenfelder, B.~H. McMahon, and F.~G. Parak, ``Slaving:
  solvent fluctuations dominate protein dynamics and functions,'' {\em Proc.
  Natl. Acad. Sci. U.S.A}, vol.~99, no.~25, pp.~16047--16051, 2002.

\bibitem{RN1}
Y.~Qin, L.~Wang, and D.~Zhong, ``Dynamics and mechanism of ultrafast
  water–protein interactions,'' {\em Proc. Natl. Acad. Sci. U.S.A}, vol.~113,
  no.~30, pp.~8424--8429, 2016.

\bibitem{RN69}
E.~A. Carter and J.~T. Hynes, ``Solvation dynamics for an ion pair in a polar
  solvent: Time‐dependent fluorescence and photochemical charge transfer,''
  {\em J. Chem. Phys.}, vol.~94, no.~9, pp.~5961--5979, 1991.

\bibitem{RN70}
B.~B. Laird and W.~H. Thompson, ``On the connection between gaussian statistics
  and excited-state linear response for time-dependent fluorescence,'' {\em J.
  Chem. Phys.}, vol.~126, no.~21, p.~211104, 2007.

\bibitem{RN51}
W.~L. Jorgensen and J.~Tirado-Rives, ``The opls [optimized potentials for
  liquid simulations] potential functions for proteins, energy minimizations
  for crystals of cyclic peptides and crambin,'' {\em J. Amer. Chem. Soc},
  vol.~110, no.~6, pp.~1657--1666, 1988.

\bibitem{RN81}
S.~Pal, P.~K. Maiti, B.~Bagchi, and J.~T. Hynes, ``Multiple time scales in
  solvation dynamics of dna in aqueous solution: the role of water,
  counterions, and cross-correlations,'' {\em J. Phys. Chem. B}, vol.~110,
  no.~51, pp.~26396--26402, 2006.

\bibitem{RN88}
K.~Furse and S.~Corcelli, ``Molecular dynamics simulations of dna solvation
  dynamics,'' {\em J. Phys. Chem. Lett.}, vol.~1, no.~12, pp.~1813--1820, 2010.

\bibitem{RN74}
M.~L. Connolly, ``Solvent-accessible surfaces of proteins and nucleic acids,''
  {\em Science}, vol.~221, no.~4612, pp.~709--713, 1983.

\bibitem{RN23}
K.~Modig, E.~Liepinsh, G.~Otting, and B.~Halle, ``Dynamics of protein and
  peptide hydration,'' {\em J. Amer. Chem. Soc}, vol.~126, no.~1, pp.~102--114,
  2004.

\bibitem{RN24}
B.~Halle, ``Protein hydration dynamics in solution: a critical survey,'' {\em
  Philosophical Transactions of the Royal Society of London B: Biological
  Sciences}, vol.~359, no.~1448, pp.~1207--1224, 2004.

\bibitem{RN25}
V.~P. Denisov and B.~Halle, ``Protein hydration dynamics in aqueous solution,''
  {\em Faraday Discuss.}, vol.~103, pp.~227--244, 1996.

\bibitem{RN87}
B.~Hess, C.~Kutzner, D.~Van Der~Spoel, and E.~Lindahl, ``Gromacs 4: algorithms
  for highly efficient, load-balanced, and scalable molecular simulation,''
  {\em J. Chem. Theo. Comp.}, vol.~4, no.~3, pp.~435--447, 2008.

\bibitem{RN53}
S.~Nos\'{e}, ``A molecular dynamics method for simulations in the canonical
  ensemble,'' {\em Mol. phys.}, vol.~52, no.~2, pp.~255--268, 1984.

\bibitem{RN54}
S.~Nos\'{e}, ``A unified formulation of the constant temperature molecular
  dynamics methods,'' {\em J. Chem. Phys.}, vol.~81, no.~1, pp.~511--519, 1984.

\bibitem{RN55}
W.~G. Hoover, ``Canonical dynamics: equilibrium phase-space distributions,''
  {\em Phys. Rev. A}, vol.~31, no.~3, p.~1695, 1985.

\bibitem{RN78}
M.~Parrinello and A.~Rahman, ``Polymorphic transitions in single crystals: A
  new molecular dynamics method,'' {\em J. Appl. Phys.}, vol.~52, no.~12,
  pp.~7182--7190, 1981.

\bibitem{RN76}
B.~Hess, H.~Bekker, H.~J. Berendsen, and J.~G. Fraaije, ``Lincs: a linear
  constraint solver for molecular simulations,'' {\em J. Comp. Chem.}, vol.~18,
  no.~12, pp.~1463--1472, 1997.

\bibitem{RN77}
T.~Darden, D.~York, and L.~Pedersen, ``Particle mesh ewald: An n⋅ log (n)
  method for ewald sums in large systems,'' {\em J. Chem. Phys.}, vol.~98,
  no.~12, pp.~10089--10092, 1993.

\end{thebibliography}
\bibliographystyle{ieeetr}

\end{document}